\documentclass[a4paper,twocolumn,nofootinbib,showpacs,aps,floatfix,superscriptaddress]{revtex4}
\usepackage{graphicx}
\usepackage{bm}
\usepackage{amsmath}
\newcommand{\ignore}[1]{}
\def\beq{\begin{equation}}
\def\eeq{\end{equation}}
\newcommand{\beqa}{\begin{eqnarray}}
\newcommand{\eeqa}{\end{eqnarray}}

\def\d{{\rm d}}

\def\om{\omega}

\newcommand {\fsin} [1] {\sin \left( #1 \right)}
\newcommand {\fcos} [1] {\cos \left( #1 \right)}
\newcommand {\fcot} [1] {\mbox{cot} \left( #1 \right)}
\renewcommand {\Re} [1] {\mbox{Re} \left( #1 \right)}
\newcommand {\ferfi} [1] {\mbox{erfi} \left( #1 \right)}
\newcommand {\fabsq}[1] {\left| #1 \right|^2}
\newcommand {\fabs}[1] {\left| #1 \right|}

\begin{document}
\title{Momentum-space interferometry with trapped ultracold atoms} 
\author{A. Ruschhaupt}
\email{a.ruschhaupt@tu-bs.de}
\affiliation{Institut f\"ur Mathematische Physik, TU Braunschweig, Mendelssohnstrasse 3, 38106 Braunschweig, Germany}
\author{A. del Campo}
\email{a.del-campo@imperial.ac.uk}
\affiliation{Institute for Mathematical Sciences, Imperial College London, SW7 2PE, UK;\\
QOLS, The Blackett Laboratory, Imperial College London, Prince Consort Road, SW7 2BW,UK}

\author{J. G. Muga}
\email{jg.muga@ehu.es}
\affiliation{Departamento de Qu\'\i mica-F\'\i sica, Universidad del Pa\'\i s Vasco, Apartado 644, 48080 Bilbao, Spain}
\begin{abstract}
Quantum interferometers are generally set so that phase differences between 
paths in coordinate space combine constructive or destructively.  
Indeed, the interfering paths can also meet in momentum space leading   
to momentum-space fringes. 
We propose and analyze a method to produce interference
in momentum space by phase-imprinting part of a trapped atomic cloud with a
detuned laser.
For one-particle wave functions analytical expressions are found for the
fringe width and shift versus the phase imprinted. The effects of 
unsharpness or displacement of the phase jump are also studied, as well as 
many-body effects  to determine the potential
applicability of momentum-space interferometry.
\end{abstract}
\maketitle
%
%
%
\section{Introduction\label{sec1}}
\begin{figure}[t]
\begin{center}
\includegraphics[angle=0,width=0.90\linewidth]{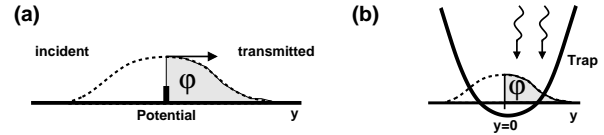}
\end{center}
\caption{\label{fig1} Schematic settings: (a) phase imprinting 
caused by wave packet passage above a weak, narrow potential, (b) phase imprinting 
caused by illumination with a detuned laser of a trapped wave packet.}
\end{figure}

In most quantum interferometers the phase differential between 
paths that meet in a coordinate-space point or region at a given time
lead to constructive or destructive wave combinations and thus 
to fringes, but    
the paths can also interfere in momentum space and produce     
momentum-space fringes. 
In particular, during the crossing of a wavepacket    
over a small and thin barrier
(compared to the energy and width of the wavepacket),  
see Fig. \ref{fig1}a,
the momentum distribution
can change dramatically, vanishing at the center of the distribution, 
and being enhanced at the wings. This process 
would violate {\em classical} energy-conservation \cite{PRL98}, 
and is due to interference in momentum space between incident and
transmitted parts of the wave \cite{PRA01,PRA05}.  

The experimental implementation of this effect is 
in principle possible
with current cold-atom technology, by turning off an effective 
detuned-laser barrier in the midst of the wavepacket passage, 
but a version which is simpler to implement is described here. 
The effect of the scattering barrier in the original proposal is to imprint an
appropriate phase on approximately half 
of the wavepacket, and this can also be achieved by shining  
part of a trapped, initially stationary, wavepacket with a strong 
laser pulse, during a short time in the scale in which a perturbation propagates (the
correlation time \cite{BS04}), see Fig. \ref{fig1}b.
The phase imprinting technique was first introduced with the purpose of
generating vortices \cite{Lewenstein99,Cornell01}. 
Here we shall study the properties of the resulting fringes, 
and show that the effect on the momentum distribution is similar to the 
effect of the scattering process, with the creation of a vanishing point (``dark notch'') 
at the center and enhancement 
of the wings. Furthermore, we shall study 
the shift, width and visibility of this central ``dark notch''
as a function of the imprinted phase, as a necessary  step to
determine the potential applicability of momentum-space interferometry. 
The imprinted phase carries information about the laser 
interaction (time, laser intensity, frequency) that can be obtained from the
notch.
By immediately removing the external trap, the momentum distribution is essentially frozen
after the imprinting, 
and many-body effects cease to play a role. 
Then, the momentum-space notch will become by expansion a 
coordinate-space notch measurable  with standard time-of-flight
techniques. Alternatively,
the momentum distribution can be accessed by stimulated Raman transitions
\cite{GW01} or through the single-particle reduced density matrix 
\cite{Bloch}.

In the following section we will describe the setting in more
detail for a single particle or
many non-interacting particles. In Section \ref{sec3} we will
consider the role of interactions within the mean-field regime
and in Section \ref{sec4} we will look at Tonks-Girardeau and
non-interacting Fermi gases.

%

%
%

\section{Non-Interacting Regime\label{sec2}}
We start considering the non-interacting regime in which the single-particle
description is valid. A highly anisotropic three-dimensional harmonic trap is assumed, 
so that the transverse degrees of freedom remain frozen and the system becomes 
effectively one-dimensional, along the axis with lowest trap frequency $\omega$.
It is useful to introduce dimensionless variables,
namely a dimensionless position $y = \sqrt{\frac{m\omega}{\hbar}} x$,
where $x$ is the dimensional position and $m$ the mass of the single particle,
and a dimensionless momentum $q = \sqrt{\frac{\hbar}{m\omega}} k$,
where $k$ is the dimensional wavenumber. 
The Hamiltonian describing the system is
$H = - \frac{\hbar^2}{2m} \frac{\partial^2}{\partial x^2}
+ \frac{m\om^2}{2} x^2$
and in the above dimensionless variables we get
$H = \frac{\hbar\omega}{2}\left(-\frac{\partial^2}{\partial y^2} + y^2\right)$.
The corresponding eigenvalues are
$E_n = \frac{\hbar\omega}{2}(2n+1)$ and
the eigenstates are $\psi_n (y) = \frac{1}{\sqrt{2^n n! \sqrt{\pi}}} H_n (y)
e^{-y^2/2}$,
where $H_n (y)$ is the $n$-th Hermite polynomial. The eigenstates are normalized
such that $\int dy\, \psi_n (y) \psi_m (y) = \delta_{n,m}$.

Initially, the trapped particle is  described by the
wavefunction $\psi_0 (y)$ in coordinate space.
Then a phase is imprinted on the
right hand side of the trap, i.e. for $y>0$. For atoms, this can be achieved by
shining an appropriate detuned laser pulse for a short time $t$.
The detuned laser acts as a mechanical potential $V \Theta(y)$
on the atom, where $V=\Omega^2\hbar/4\Delta$, $\Omega$ is the Rabi frequency, and $\Delta$ 
the detuning (laser frequency minus transition frequency). If the time
$t$ is short, the effect is to imprint a phase
$\varphi=-Vt/\hbar$ on the wave function for $y>0$. The
wavefunction in coordinate space becomes 
$\psi_0 (y) e^{i\varphi w(y)}$ with $w(y) = \Theta(y)$, 
and in momentum space 
\begin{eqnarray*}
\phi_0 (q) &=& \frac{1}{\sqrt{2\pi}} \int_{-\infty}^0 dy\, \psi_0 (y) e^{-i q y}
\\
&+& e^{i\varphi} \frac{1}{\sqrt{2\pi}} \int_{0}^{\infty} dy\, \psi_0 (y) 
e^{-i q y}.
\end{eqnarray*}
Each momentum gets an amplitude contribution from two different terms 
and we may expect interferences in $\fabsq{\phi_0 (q)}$ for $\varphi > 0$.
In the following, this interference pattern will be studied
and analytical expressions will be found 
for the fringe shift, width and visibility versus the phase $\varphi$ imprinted.
The effects of unsharpness or spatial displacement of the phase
jump are also studied.

\subsection{Reference Case}
Let us first study the effect of imprinting
a phase $\varphi$ on the ground state of the harmonic trap, $n=0$.
In this case the momentum probability density becomes
\begin{eqnarray}
\fabsq{\phi_0 (q)}=\frac{e^{-q^2}}{\sqrt{\pi}}
\fabsq{\fcos{\frac{\varphi}{2}} + \fsin{\frac{\varphi}{2}}
\ferfi{\frac{q}{\sqrt{2}}}}, 
\label{wave00}
\end{eqnarray}
which has a zero at $q_0$, a solution of
\begin{eqnarray}
\ferfi{\frac{q_0}{\sqrt{2}}} = -\fcot{\frac{\varphi}{2}}.
\label{eq1}
\end{eqnarray}

\begin{figure}[t]
\begin{center}
\includegraphics[angle=0,width=\linewidth]{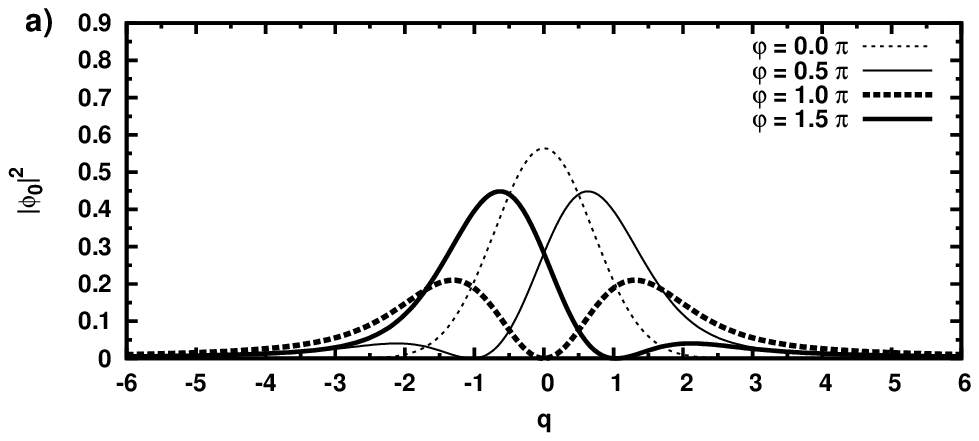}

\includegraphics[angle=0,width=\linewidth]{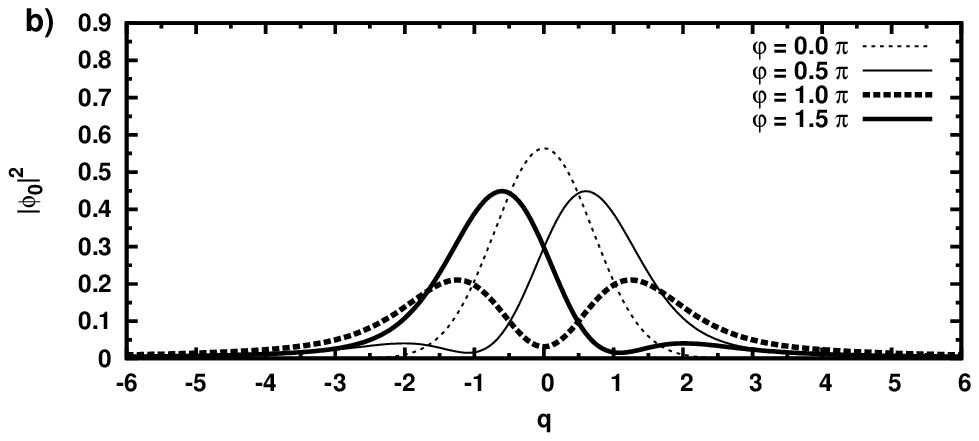}

\includegraphics[angle=0,width=\linewidth]{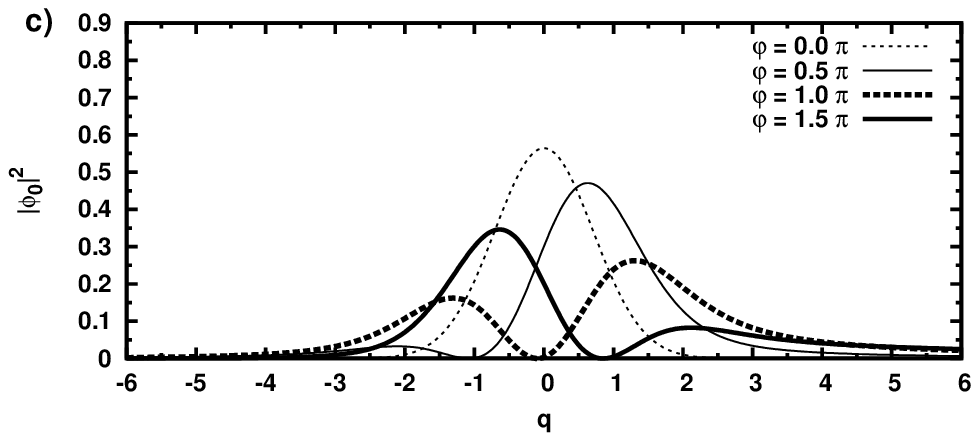}

\includegraphics[angle=0,width=\linewidth]{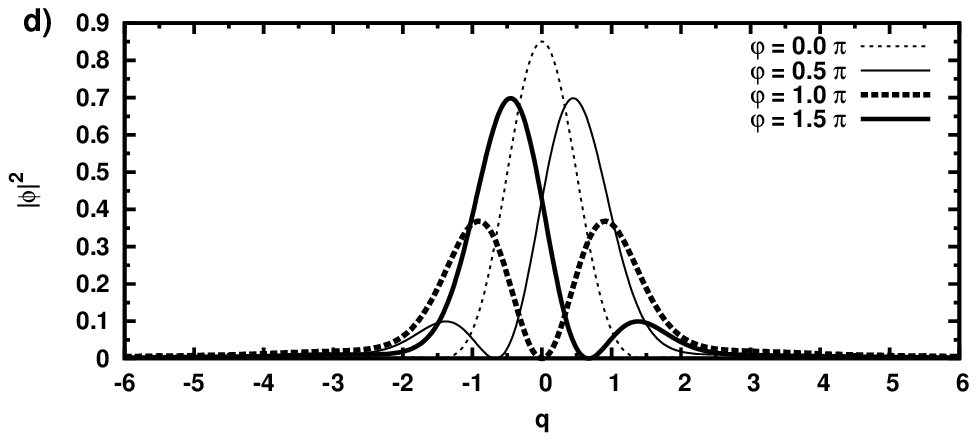}
\end{center}
\caption{\label{fig2} Wave function in momentum space
for different phase imprintings $\varphi$;
(a) reference case, $y_0 = 0$, $\zeta = 0$,
(b) effect of shifting, $y_0 = 0.3$, $\zeta = 0$,
(c) effect of smoothing, $y_0 = 0$, $\zeta = 0.1$,
(d) effect of interaction, solution of the GPE:
$y_0=0$, $\zeta = 0$, $g = 20$.}
\end{figure}

\begin{figure}[t]
\begin{center}
\includegraphics[angle=0,width=\linewidth]{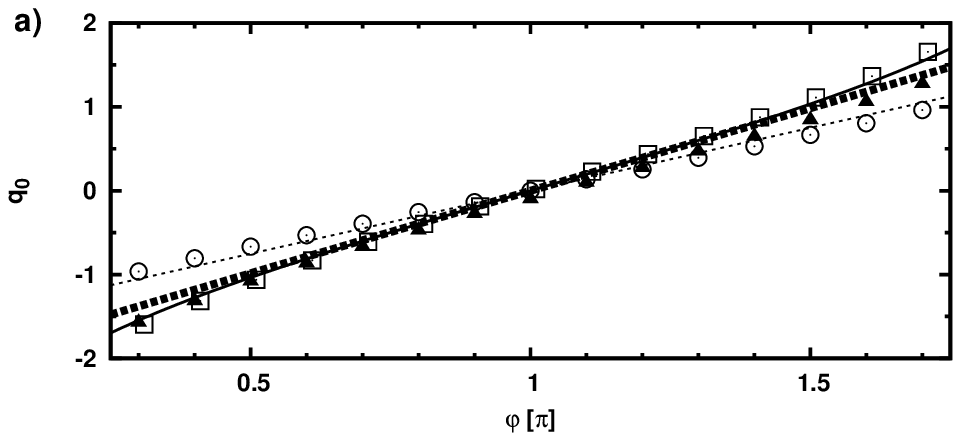}

\includegraphics[angle=0,width=\linewidth]{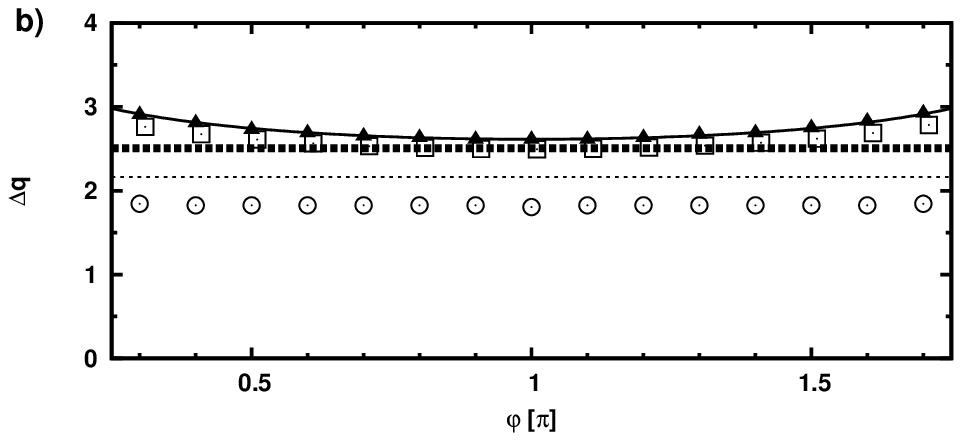}

\includegraphics[angle=0,width=\linewidth]{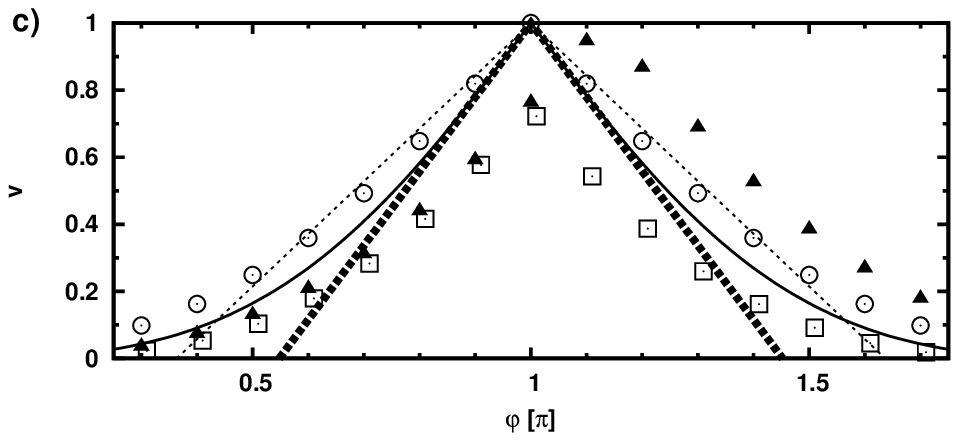}
\end{center}
\caption{\label{fig3}
(a) Momentum of the minimum versus $\varphi$,
(b) width of the minimum versus $\varphi$,
(c) visibility of the minimum versus $\varphi$;
in all cases $n=0$.\\
Analytical approximations: 
thick dotted lines:
$\tilde{v}$, $\widetilde{\Delta q}$, $\tilde{q}_0$;
dotted lines: $\bar{v}$, $\overline{\Delta q}$, $\bar{q}_0$ (Thomas-Fermi with
$g=20$).\\
Exact results:
solid line: reference case, $v$, $\Delta q$, $q_0$ with $y_0 = 0, \zeta = 0$;
boxes: effect of shifting, $v$, $\Delta q$, $q_0$ with $y_0 = 0.3, \zeta = 0$;
triangles: effect of smoothing, $v$, $\Delta q$, $q_0$ with $y_0 = 0, \zeta = 0.1$;
circles: effect of interaction, $v$, $\Delta q$, $q_0$ based on the solution of the GPE with
$y_0 = 0, \zeta = 0, g = 20$.}
\end{figure}

Momentum distributions for this reference case after different phase
imprintings are displayed in Fig. \ref{fig2}a.
Note the optimality of $\varphi=\pi$ to produce 
a deep minimum, in fact a zero, exactly at the peak of the original
distribution, $q_0=0$, and the enhancement at the wings.
In the following, we will concentrate on this central ``dark notch''.
The ``motion'' of $q_0$ with $\varphi$ can be seen in Fig. \ref{fig2}a, and in more 
detail in Fig. \ref{fig3}a (solid line),  
where $q_0$ is plotted versus $\varphi$.
It shows a linear behavior in $\pi/2 < \varphi <
3\pi/2$ and slight deviations beyond that range. The width $\Delta q=q_+ - q_-$ of the central
interference dip
is defined as the difference between the momentum $q_+$ of the maximum on the
right-hand side of the minimum and the momentum $q_-$ of the
maximum on the left-hand side.
Fig. \ref{fig3}b (solid line) shows the width versus the phase.
Note that the width is always greater that $\sqrt{2\pi}$ (value
of the thick dotted line).

Another important quantity is the visibility of the minimum, which 
we define as
\begin{eqnarray*}
v = \frac{\min_{\pm} \left(\fabsq{\phi_0 (q_\pm)} - \fabsq{\phi_0 (q_0)}\right)}{
\frac{\fabsq{\phi_0 (q_+)} +  \fabsq{\phi_0 (q_-)}}{2} + \fabsq{\phi_0 (q_0)}}.
\end{eqnarray*}
This visibility is plotted in Fig. \ref{fig3}c (solid line).
From the calculations we can infer that 
the visibility limits the working range of the interferometer to 
$\pi/2<\phi<3\pi/2$, and is optimal around $\phi=\pi$.  

%

%

\subsection{Analytical approximations for the reference case\label{sec2approx}}

The goal now is to derive approximate analytical formulae
describing the properties of the central ``dark notch'' as a function of the imprinted phase. 
From $\frac{\partial \fabsq{\phi_0 (q)}}{\partial q} = 0$, 
we get the extreme points of $\fabsq{\phi_0 (q)}$ as solutions of
\begin{eqnarray}
&&\overbrace{\left(\fcos{\frac{\varphi}{2}} + \fsin{\frac{\varphi}{2}}
\ferfi{\frac{q}{\sqrt{2}}}\right)}^{z_1}
\nonumber\\
&\times& 
\underbrace{\left(q \fcos{\frac{\varphi}{2}} - \fsin{\frac{\varphi}{2}}
\left(e^{q^2/2}\sqrt{\frac{2}{\pi}} - q\, \ferfi{\frac{q}{\sqrt{2}}}\right)
\right)}_{z_2}
\nonumber\\
&=&0.
\label{eqy}
\end{eqnarray}
Note that if $q$ is a solution of (\ref{eqy}) for 
$\varphi = \pi + \Delta\varphi$ then $-q$ is a solution for
$\varphi = \pi - \Delta\varphi$.

One of the solutions of equation (\ref{eqy}) fulfilling $z_1 = 0$
and describing the momentum of the minimum is approximately given by
\begin{eqnarray}
q_0 \approx \sqrt{\frac{\pi}{2}} \frac{\varphi-\pi}{2} =: \tilde q_0
\label{q0approx}
\end{eqnarray}
(based on a linearization
of $z_1 = 0$ around $\varphi \approx \pi$ and $q \approx 0$).
This describes a linear displacement of the minimum with $\varphi$. 
Fig. \ref{fig3}a shows the exact momentum of the minimum $q_0$ (solid line)
and $\tilde q_0$ (thick dotted line) versus $\varphi$. 

Now we shall obtain expressions for the (momentum of the)
left maximum, $q_-$,
and the right maximum $q_+$.
An approximate solution of $z_2 = 0$ is
\begin{eqnarray*}
q_+ \approx \frac{\varphi}{\sqrt{2\pi}} =: \tilde q_+,
\end{eqnarray*}
which follows from a linearization
of $z_2 = 0$ around $\varphi \approx 0$ and $q \approx 0$.
Another approximate
solution of $z_2 = 0$ is
\begin{eqnarray*}
q_- \approx -\frac{2\pi - \varphi}{\sqrt{2\pi}} =: \tilde q_-,
\end{eqnarray*}
obtained by linearizing $z_2 = 0$ around $\varphi \approx 2\pi$
and $q \approx 0$.
Thus we get for the 
the width of the interference dip
\begin{eqnarray*}
\Delta q = q_+ - q_- \approx \tilde{q}_+ - \tilde{q}_- = \sqrt{2\pi}
=: \widetilde{\Delta q}.
\end{eqnarray*}
Figure \ref{fig3}b compares the numerically calculated exact width
$\Delta q$ (solid line) with $\widetilde{\Delta q}$
(thick dotted line).

We can also find a simple expression for the visibility. 
From Eq. (\ref{wave00}) and using also the
approximations of the momentum of the two maxima and
the minima and retaining only the first order in $\varphi$ we get
\begin{eqnarray*}
\fabsq{\phi_0 (q_-)} &\approx& \alpha + \beta (\varphi - \pi),\\
\fabsq{\phi_0 (q_0)} &\approx& 0,\\
\fabsq{\phi_0 (q_+)} &\approx& \alpha - \beta (\varphi - \pi),\\
\end{eqnarray*}
where
\begin{eqnarray*}
\alpha &=&  \frac{e^{-\pi/2}}{\sqrt{\pi}} \ferfi{\frac{\sqrt{\pi}}{2}}^2
\approx 0.210,\\
\beta &=& \frac{e^{-\pi/2}}{\pi^{3/2}} \ferfi{\frac{\sqrt{\pi}}{2}}
\left(-2 e^{\pi/4} + \pi + \pi \ferfi{\frac{\sqrt{\pi}}{2}}\right)\\
&\approx& 0.148.
\end{eqnarray*}
The final result is 
\begin{eqnarray*}
v \approx 1 - \frac{\beta}{\alpha} \fabs{\varphi - \pi} =: \tilde{v},
\end{eqnarray*}
also shown in Fig. \ref{fig3}c (thick dotted line).
It gives a lower bound
for the exact result $v$ (solid line).

\subsection{Perturbations of the reference case}

In this subsection we examine the effect of perturbations
of the reference case.
First we want to discuss the effect of shifting the edge 
of the phase imprinted region, $y_0$, out of the center of the trap,
i.e. we have $w(z) = \Theta (y-y_0)$.
Right after the phase-imprinting,
the $0$-th eigenstate now becomes
$\psi_0(y)e^{i\varphi w(z)}=
\psi_0(y)e^{i\varphi\Theta(y-y_0)}$,
and the momentum distribution becomes 
\begin{eqnarray}
\fabsq{\phi_0 (q)}=\frac{e^{-q^2}}{\sqrt{\pi}}
\fabsq{\fcos{\frac{\varphi}{2}} + \fsin{\frac{\varphi}{2}}
\ferfi{\frac{q-i y_0}{\sqrt{2}}}}.
\label{wave0}
\end{eqnarray}
From 
$\frac{\partial \fabsq{\phi_0 (q)}}{\partial q} = 0$, 
we get the extreme points as solutions of
\begin{eqnarray}
 \sqrt{\frac{\pi}{2}} q \exp[-\frac{1}{2} (q^2 - y_0^2)]\fabsq{z} &=&
\fsin{\frac{\varphi}{2}} \Re{z e^{i q y_0}},\nonumber\\
\label{eq_exact}
\end{eqnarray}
where $z = \fcos{\frac{\varphi}{2}} + \fsin{\frac{\varphi}{2}}
\ferfi{\frac{q-i y_0}{\sqrt{2}}}$.
Again, if $q$ is a solution of (\ref{eq_exact}) for 
$\varphi = \pi + \Delta\varphi$ then $-q$ is a solution for
$\varphi = \pi - \Delta\varphi$. In addition, because (\ref{eq_exact})
is not changing if $y_0$ is replaced by $-y_0$, the solutions are
the same for $\pm y_0$.
If $y_0 \ll1$, to first order in $y_0$ Eq. (\ref{wave0})
and Eq. (\ref{eq_exact}) are independent of $y_0$
and therefore the above derived approximations in Section \ref{sec2approx}
for the case $y_0 = 0$ still hold.
An example for $y_0 = 0.3$ is plotted in Fig. \ref{fig2}b.
The corresponding momentum, width, and visibility of the minima versus
$\varphi$ is also plotted in Fig. \ref{fig3} (boxes).
The main effect of increasing $y_0$ is to lower the visibility
(Fig. \ref{fig3}c).

Finally, we want to look at the effect of a more realistic smooth 
profile of the imprinted phase, instead of using an 
idealized step function. Therefore, we consider now a sigmoid function 
\beqa 
w(y)=\frac{1}{2}[1+{\rm tanh}(y/\zeta)],
\label{sigmoid}
\eeqa 
which for $\zeta\rightarrow0$ becomes $\Theta(y)$.
The results for a smoothing $\zeta = 0.1$ can be seen in Fig \ref{fig2}c
and in Fig. \ref{fig3} (triangles). Smoothing
results mainly in a shift of the maximum of the visibility
(see Fig. \ref{fig3}c).

%
%

\subsection{Momentum interference for Excited States}
We shall next consider the effect of phase imprinting on
excited states of the
harmonic trap with the simplest profile $w(y)=\Theta(y)$. 
The probability amplitude in momentum space is then given by
\begin{eqnarray*}
& & \fabsq{\phi_n (q)} = \frac{1}{2\pi 2^n n! \sqrt{\pi}}\nonumber\\
& & \times
\fabsq{\int_0^{\infty} dy\, H_n (y) e^{-y^2/2}
\left[ (-1)^n e^{i y q} + e^{i\varphi} e^{- i y q}\right]},
\end{eqnarray*}
which clearly simplifies for $n=0$ to Eq. (\ref{wave00}).

Let us look for $q_0$ fulfilling $\fabsq{\phi_n (q_0)} = 0$,
i.e. for the momentum of the minimum.
Assuming $q_0 \ll 1$ such that
$(-1)^n e^{i y q_0} + e^{i\varphi} e^{- i y q_0}
\approx ((-1)^n + e^{i \varphi}) + i q_0 ((-1)^n - e^{i\varphi}) y$, 
%
\begin{eqnarray*}
\fabsq{\phi_n (q_0)}
&\approx&\frac{1}{2\pi 2^n n! \sqrt{\pi}}
\nonumber\\
&\times&
\fabsq{
((-1)^n + e^{i \varphi}) 
A_n
+ i q_0 ((-1)^n - e^{i\varphi})
B_n},
\end{eqnarray*}
where we have introduced $A_n=\int_0^{\infty} dy\, H_n (y) e^{-y^2/2}$, 
and $B_n=\int_0^{\infty} dy\, y H_n (y) e^{-y^2/2}$.
Solving this for $\fabsq{\phi_n (q_0)} = 0$, we get
\begin{eqnarray}
q_0 \approx i \frac{(-1)^n +  e^{i \varphi}}{(-1)^n - e^{i\varphi}}
\frac{A_n}{B_n}.
\label{eq_q0}
\end{eqnarray}
The cases in which $n$ is even or odd will be examined separately.

\begin{figure}[t]
\begin{center}
\includegraphics[angle=0,width=\linewidth]{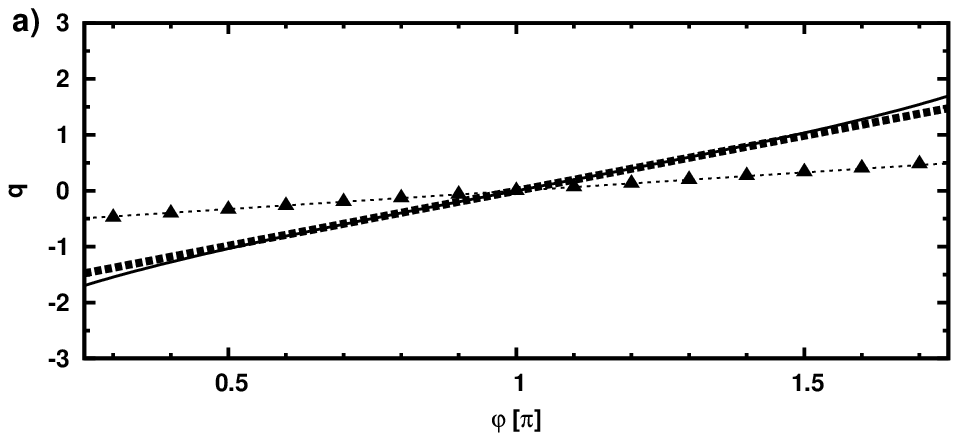}

\includegraphics[angle=0,width=\linewidth]{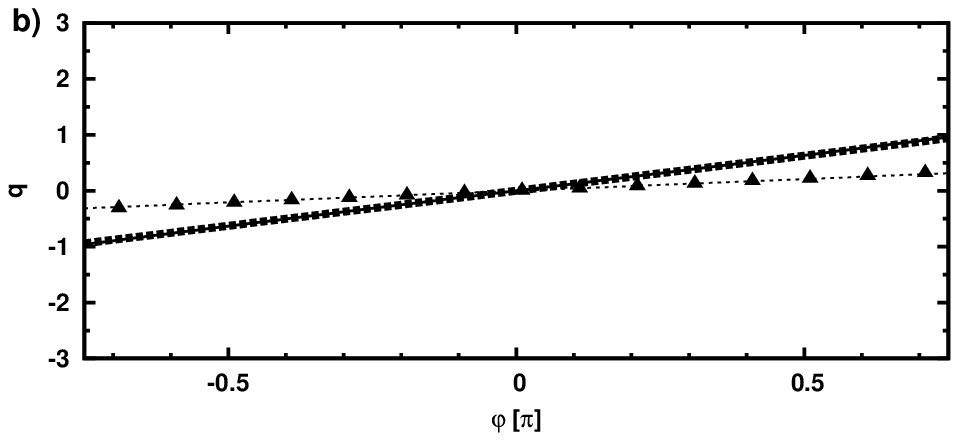}

\includegraphics[angle=0,width=\linewidth]{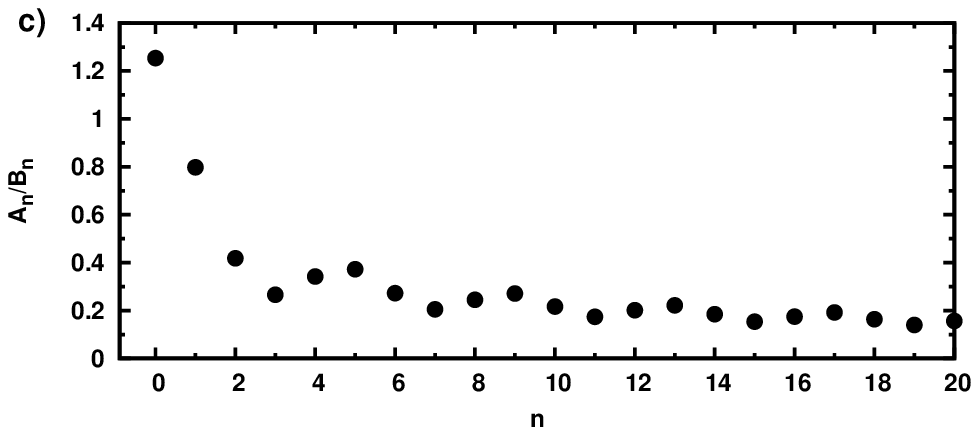}
\end{center}
\caption{\label{fig4} Effect for excited states:
(a,b) Momentum of minimum versus $\varphi$; 
(a) $n=0$: $q_0$ (solid line),
$\tilde{q}_0$ (thick dotted line);
$n=2$: $q_0$ (triangles),
$\tilde{q}_0$ (dotted line);
(b) $n=1$: $q_0$ (solid line),
$\tilde{q}_0$ (thick dotted line);
$n=3$: $q_0$ (triangles),
$\tilde{q}_0$ (dotted line);
(c) Ratio $\frac{A_n}{B_n}$ versus $n$. 
}
\end{figure}

\paragraph{$n$ even:}
We are interested in the motion of the
zero $\fabsq{\phi_n (q_0)} =0$ for $\varphi \approx \pi$.
From Eq. (\ref{eq_q0}) we get in first order in $\varphi-\pi$ that
\begin{eqnarray*}
q_0 \approx \frac{A_n}{B_n} \frac{\varphi-\pi}{2} =: \tilde{q}_0.
\end{eqnarray*}
Examples for the exact solution $q_0$ and the approximation $\tilde{q}_0$
for $n=0,2$ can be found in Figure \ref{fig4}a.

\paragraph{$n$ odd:}
Now we are interested in the motion of
the zero $\fabsq{\phi_n (q_0)} =0$ versus $\varphi$ for $\varphi \approx 0$. 
From Eq. (\ref{eq_q0}) we get in first order in $\varphi$ that
\begin{eqnarray*}
q_0 \approx \frac{A_n}{B_n} \frac{\varphi}{2} =: \tilde{q}_0.
\end{eqnarray*} 

Examples for the exact solution $q_0$ and the approximation $\tilde{q}_0$
for $n=1,3$ can be found in Figure \ref{fig4}b.

In addition, Fig. \ref{fig4}c shows the value of the ratio $\frac{A_n}{B_n}$
for odd and even $n$. Clearly, increasing $n$ makes the interferometer
less sensitive to phase variations and $n=0$ provides the optimal behavior.  

%

%
%
%
\section{The mean-field regime\label{sec3}}
We shall consider now the role of the interactions within the mean-field approach. 
In the mean-field regime, for low enough temperatures the phase fluctuations can be suppressed \cite{GanShl03}.
Weakly interacting ultracold gases in 1D are then described by the
Gross-Pitaevskii equation (GPE).

\begin{figure}[t]
\begin{center}
\includegraphics[angle=0,width=\linewidth]{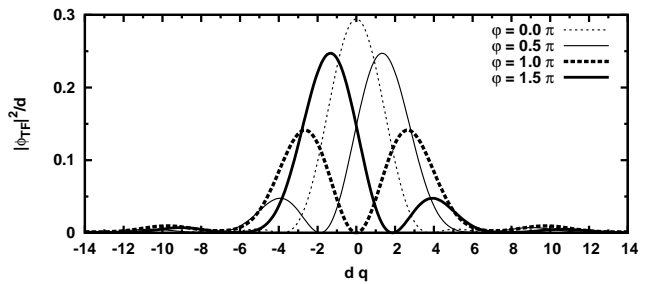}
\end{center}
\caption{\label{fig5} Interference in the momentum distribution of a Bose-Einstein condensate in the Thomas-Fermi regime.}
\end{figure}
%

Assume that an effectively 1D Bose-Einstein condensate
is prepared in the harmonic trap. The condensate wave function is the
ground state of the 1D (stationary) GPE
\begin{equation*}
\mu \Psi(x) 
=- \frac{\hbar^2}{2m}
\frac{\partial^2\Psi(x)}{\partial x^2}  + \frac{m\omega^2}{2} x^2\nonumber\\
+\, \frac{\hbar}{2} g_{1D} \fabsq{\Psi (x)} \Psi(x),
\end{equation*}
where $\mu$ is the chemical potential and $g_{1D}$ the effective 1D coupling
parameter related to the three-dimensional scattering length
\cite{Olshanii98}. We assume that $\int dx \fabsq{\Psi(x)} = 1$.
By introducing $u := 2\mu/(\hbar\omega)$, $g :=
\sqrt{\frac{m}{\hbar\omega}} g_{1D}$, and 
$\psi (y) := \sqrt[4]{\frac{\hbar}{m\omega}} 
\Psi \left(\sqrt{\frac{\hbar}{m\omega}} y \right)$
we can write this equation in dimensionless form, 
\begin{equation}
u \psi(y) 
=- \frac{\partial^2\psi(y)}{\partial y^2}  + y^2\nonumber\\
+\, g \fabsq{\psi (y)} \psi(y),
\label{gp}
\end{equation}
with $\int dy \fabsq{\psi(y)} = 1$.
The ground state can be numerically computed by
using the imaginary time method \cite{gr14, gr16}.
For $\varphi=0$, it is well known that as the mean-field
interaction is increased, 
the density profile becomes more uniform, 
while the resulting momentum distribution $\fabsq{\phi}$
is sharply peaked \cite{BP96,DPS96}.
To study the effect of a small $g$
as a perturbation of the previous results, 
we shall  
imprint a phase $\varphi$ on the ground state wavefunction and calculate the
minimum of the resulting interference pattern in momentum-space.
An example with $g=20$ is shown in Fig. \ref{fig2}d.
The visibility, the width and the momentum of the minimum versus $\varphi$
with $g=20$ is shown in Fig. \ref{fig3} (circles).
The main effect is that the
slope in Fig. \ref{fig3}b decreases with increasing atom-atom
interaction, i.e. the sensitivity of the interferometer with
respect to $\varphi$ decreases with increasing $g$. A 
large atom-atom interaction $g$ may also perturb    
the measurement of the momentum distribution
by time-of-flight techniques.
There is however also a positive effect: an increase of $g$ 
makes the interference dip sharper and
improves the visibility,  
see Fig. \ref{fig3}c.

It is possible to derive analytical approximate formulae for large $g$.
For $g\gg1$ the condensate enters into the Thomas-Fermi regime
\cite{BP96,DPS96}. The mean-field interaction is then so large that the
kinetic energy can be neglected in the Hamiltonian so that the
time-independent GPE reads 
$u \psi(y)= \left(y^2+ g \fabsq{\psi (y)}\right) \psi(y)$.
The Thomas-Fermi wavefunction is then given by
$\psi_{TF}(y)=\sqrt{(u-y^2)/g}$ with $u=(3g/4)^{2/3}$ whenever $\fabs{y} <
d$ and zero elsewhere;
$d = \sqrt{u}$ is the Thomas-Fermi half-width.

The probability distribution in momentum space after a phase imprinting
$\varphi$ with a profile $w(y) = \Theta(y)$ is given in this case by
\beqa
|\phi_{TF}(q)|^2 = \frac{3 \pi}{8d q^2}
\left[J_{1}(qd)\fcos{\frac{\varphi}{2}}
+{\bf H}_1(qd)\fsin{\frac{\varphi}{2}}\right]^2.
\eeqa
Here, $J_1(k)$ is the Bessel function of first order and ${\bf H}_1(y)$ is
the first order Struve function \cite{AS65}.
$|\phi_{TF}(q)|^2/d$ is plotted for different
values of $\varphi$ in Fig. \ref{fig5}.
Again there is a minimum for $\varphi = \pi$ at $q=0$ which
is shifted if $\varphi$ is changed.

The minima and the maxima of $|\phi_{TF}(q)|^2$ for a fixed $\varphi$
can be found by looking at the zeros of the derivative, this leads
to the equation
\begin{eqnarray*}
\underbrace{\frac{1}{qd}\left[J_{1}(qd)\fcos{\frac{\varphi}{2}}
+{\bf H}_1(qd)\fsin{\frac{\varphi}{2}}\right]}_{a_1} \times & &\\
\underbrace{\frac{1}{(qd)^2}\left[\pi qd J_2(qd) \fcos{\frac{\varphi}{2}}
+ \left(2+\pi qd {\bf H}_{-2}(qd)\fsin{\frac{\varphi}{2}}\right)\right]
}_{a_2} & &\\
= 0. & &
\end{eqnarray*}
Making $a_1 = 0$ and using a linearization around $q \approx 0$
and $\varphi \approx \pi$, we arrive at
\begin{eqnarray*}
q \approx \frac{3\pi}{8d} (\varphi - \pi) =: \bar{q}_0,
\end{eqnarray*}
where $d = (3g/4)^{1/3}$, which allows to find $d$ measuring the notch
displacement. 
Making $a_2 = 0$ and using a linearization around $q \approx 0$
and $\varphi \approx 0$, we arrive at
\begin{eqnarray*}
q \approx \frac{8}{3\pi d} \varphi =: \bar{q}_+,
\end{eqnarray*}
and by using a linearization around $q \approx 0$
and $\varphi \approx 2\pi$, we arrive at
\begin{eqnarray*}
q \approx -\frac{8}{3\pi d} (2\pi - \varphi) =: \bar{q}_-.
\end{eqnarray*}
An estimate of the width is then  
$\overline{\Delta q} = \bar{q}_+ - \bar{q}_- = \frac{16}{3d}$.
An approximation for the visibility can be also derived as in Section
\ref{sec2approx},
\begin{eqnarray*}
v \approx 1 - 0.5 \fabs{\varphi - \pi} =: \bar{v}.  
\end{eqnarray*}
The approximate values of the notch momentum, width, and visibility in the Thomas-Fermi regime are also plotted
in Fig. \ref{fig3} (dotted lines).

\section{The Tonks-Girardeau and non-interacting Fermi gases\label{sec4}}

At low enough densities, and under tight-transverse confinement, ultracold gases
enter the Tonks-Girardeau (TG) regime \cite{Girardeau60}, in which the
strength of the effective short-range interactions becomes so large that the
mean-field theory fails \cite{GW00}. Fortunately,
Bose-Fermi duality offers a powerful and exact approach, exploiting the
similarities between the TG and spin-polarized non-interacting Fermi gases. 
The ground-state wavefunction of the later in a harmonic trap is the familiar
Slater determinant,
$\Psi_{F}(y_{1},\dots,y_{N}) =\frac{1}{\sqrt{N!}}{\rm
  det}_{n,k=(0,1)}^{(N-1,N)}\psi_{n}(y_{k})$, built from the set of
single-particle orthonormal eigenstates $\{\psi_{n}(y)\}$.
Such atom Fock state can be efficiently prepared using the atom culling
technique as described in \cite{atomculling}. Note that the  wavefunction
$\Psi_{F}$ is totally antisymmetric and vanishes whenever the positions 
of two particles coincide.
The TG wavefunction is obtained from $\Psi_F$ by imposing the correct symmetry
under permutation of particles, i.e.,
using the Fermi-Bose (FB) mapping \cite{Girardeau60} 
\beqa
\Psi_{TG}(y_{1},\dots,y_{N})= \prod_{1\leq j<k\leq N}{\rm sgn}(y_{k}-y_{j})\Psi_{F}(y_{1},\dots,y_{N}).
\nonumber
\eeqa
Clearly, both dual systems share the same density profile \cite{GW00}
$\rho_{TG/F}(y,t)= N\!\!\int\vert\Psi_{TG/F}(y,y_{2},\dots,y_{N};t)\vert^{2} \d y_{2}
\cdots\d y_{N}
=\sum_{n=0}^{N-1}\vert\psi_{n}(y,t)\vert^{2}$, as it is the case for any
other local correlation function. However, their momentum distributions
\beqa
n(q)=(2\pi)^{-1}\int\d y\d y'e^{iq(y-y')}\rho(y,y')
\eeqa
are  drastically
different. Provided that the reduced single-particle density matrix (RSPDM) of
spin-polarized fermions is 
\beqa
\rho_F(y,y')=\sum_{n=0}^{N-1}\psi_{n}^*(y)\psi_n(y'),
\eeqa
the momentum distribution is the sum
$n_F(q)=\sum_{n=0}^{N-1}|\tilde{\psi}_n(q)|^2$ (where $\tilde{\psi}_n$ denotes the
Fourier transform of $\psi_n$).
For the TG gas, an efficient way of computing the RSPDM has been introduced
\cite{PB07,dark07}, namely, 
\beqa
\rho_{TG}(y,y')=\sum_{l,n=0}^{N-1}\psi_l^*(y){\rm
  A}_{ln}(y,y')\psi_n(y'),
\eeqa 
where ${\rm \bf A}(y,y')=({\bf P}^{-1})^T {\rm
  det}{\bf P}$ and the elements of the matrix ${\bf P}$ are
$P_{ln}(y,y')=\int\d z \psi_{l}^*(z)\phi_n(z){\rm sgn}(z-y){\rm sgn}(z-y')$, 
which reduces to $P_{ln}=\delta_{ln}-2\int_{y}^{y'}\d z \psi_{l}^*(z)\phi_n(z)$ 
for $y<y'$ without loss of generality 
. 
The momentum distribution of the TG gas, can thus be obtained as a double Fourier transform.
%
%
\begin{figure}[t]
\includegraphics[width=8.5cm,angle=0]{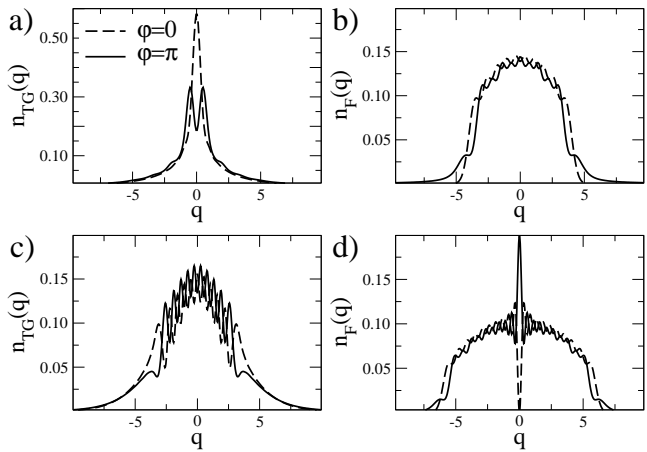}

\caption{\label{fig6} Interference in momentum space, $N=10$,
a $\varphi=\pi$ phase is imprinted for $y>0$ (solid line),
$\varphi=0$ case (dashed line);
(a) TG gas, (b) Fermi gas,
(c) TG gas after parity-selective evaporation,
(d) Fermi gas after parity-selective evaporation.}
\end{figure}
%

We consider a phase imprinting with $w(y) = \Theta(y)$.
Under this phase imprinting, a remarkable difference arises between the momentum
distribution of both dual systems. 
For a moderate $N$ the visibility of the interference fringes in the TG gas is
reduced (see Fig. \ref{fig6}a) 
but in the fermionic case the pattern has been washed out completely 
(see Fig. \ref{fig6}b).
For larger $N$ the visibility of the TG dip decreases.   
It is hence clear that the observation of such effect in any of these dual
systems would be difficult, and we turn our attention to a closely related
alternative approach.

%
\begin{figure}[t]
\includegraphics[width=8.5cm,angle=0]{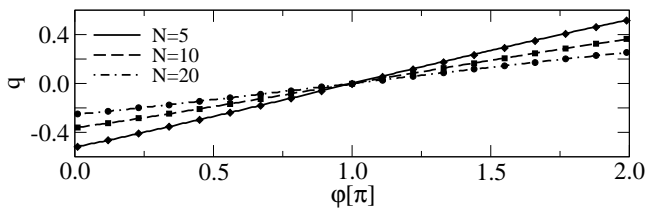}

\caption{\label{fig7} Displacement of the maximum of the momentum distribution as a function of 
the phase imprinted, for parity selective evaporation, 
fermionic cloud with only odd states. 
The symbols correspond to a smooth phase-imprinting profile with $\zeta=1/2$, whereas for the lines $\zeta=0$. 
} 
\end{figure}
%

Recently, a parity-selective evaporation (PSE)
method has been proposed which allows to prepare in principle excited states 
composed exclusively of odd-parity single-particle eigenstates
\cite{dark06}. This is achieved by shinning a  
blue-detuned laser at $y\sim0$ which removes the even-parity eigenstates. For a
spin-polarized Fermi gas 
the excited many-body wavefunction becomes $\Psi_{F}(y_{1},\dots,y_{N})
=\frac{1}{\sqrt{N!}}{\rm det}_{n,k=1}^{N}\psi_{2n-1}(y_{k})$. The
corresponding momentum distribution exhibits a 
well-defined zero at $q=0$ for all $N$ which is stationary, and robust against
significant smoothing of the phase-imprinting profile. 
The TG wavefunction equally follows from the Bose-Fermi map for PSE-prepared
states. However, the $n_{TG}(q)$ is qualitatively insensitive to the selected
parity of the single-particle states, and lacks any principal peak (or dip)
potentially useful for momentum-space interferometry (see Fig. \ref{fig6}c). 
On the other hand, the pattern $n_F(q)$ in the fermionic case
is reversed under phase imprinting, turning a zero into a peak in the momentum
distribution (see Fig. \ref{fig6}d).
Let us consider again a phase imprinting with the sigmoid profile
(\ref{sigmoid}). In Fig. \ref{fig7} we have
calculated the shift of the maximum in $n_{F}(q)$ as a function of
$\varphi$ for the cases $\zeta = 0$ (Heaviside function) and 
$\zeta= 0.5$. The dependence is found to be linear
even in the presence of the large
smoothing in the profile $\zeta=0.5$.

Therefore, between both dual systems the TG gas is preferred using phase
imprinting, whereas in combination with PSE, 
the fermionic system is a better candidate. 

\section{Discussion}

The localized phase imprinting method \cite{BS04} on trapped cold atoms
has been discussed up to now mostly 
in connection with the generation and study of solitons. 
In this paper we have instead focused on the 
characterization of the momentum distribution right after the phase 
imprinting.

First, phase imprinting of half of the wavepacket can be 
regarded as a simple way to realize the interferometry in momentum space 
that has been previously put forward for more complex scattering 
processes between cold atoms an weak laser barriers \cite{PRL98,PRA01,PRA05}.
Similar to the scattering setting, a central ``dark notch'' appears in the
momentum distribution after phase imprinting, as well as 
an enhancement of the wings. 
An advantage with respect to the scattering method 
is that there is no need to make the width
in momentum of the incident wave packet small to get 
the same
transmission coefficient and therefore the same phase
shift for all momenta. Thus we can make the trap tighter
and tighter increasing the sensitivity. 

Furthermore, the characterization of the momentum distribution
is a preliminary step to determine the potential applicability
of momentum interferometry where an unknown phase 
should be determined from the momentum shift of the central ``dark notch''.

We have studied different configurations, regimes, and perturbations.  
The momentum dark notch for non-interacting particles in the ground state
provides the most sensitive meter for the imprinted phase among the different
states considered. In dimensionless units, the momentum
shift of the ``dark notch''
versus the imprinted phase is in this case approximately given
by $\tilde q_0 = \sqrt{\frac{\pi}{2}} \frac{\varphi-\pi}{2}$ (see
Eq. (\ref{q0approx})). In dimensional units we get for the velocity
\begin{eqnarray*}
\tilde{v}_0 = 
\sqrt{\frac{\hbar\omega}{m}} \left(\frac{\pi}{2}\right)^{3/2}
\frac{\varphi-\pi}{\pi},
\end{eqnarray*}
such that we can enhance the sensitivity concerning phase differences
in principle to arbitrary high values by letting $\omega \to \infty$.
If the external trap is immediately removed
after the phase imprinting, the momentum distribution is essentially frozen
and the velocity $\tilde{v}_0$ can be measured with standard
time-of-flight technique. Assuming a free time of flight of duration $t$
after the phase imprinting
the ``dark notch'' will move a distance $\tilde{s}_0 = t \tilde{v}_0$.
If we have a spacial resolution of $\Delta s$ then the resolvance $r$
of our momentum interferometer can be defined for the reference case as
\begin{eqnarray*}
r := \frac{\pi}{\Delta\varphi} = \frac{t}{\Delta s}
\sqrt{\frac{\hbar\omega}{m}} \left(\frac{\pi}{2}\right)^{3/2}, \end{eqnarray*} 
where $\Delta\varphi$ is the minimum resolvable deviation of the phase from $\pi$. 
For $t=200 \,\mbox{ms}$, $m=$mass($^{87}$Rb), $\omega=2\pi\times 2\,\mbox{kHz}$
and $\Delta s = 5 \mu{\mbox m}$, we get
$r \approx 239$.
The effects of unsharpness or spatial displacement of the phase
jump have also been studied and the results qualitatively still holds.
Many-body effects in the mean-field regime  
lead to a mild sensitivity loss but also to an interesting increase
of visibility.
In all cases there is still a linear dependence of the
``dark notch'' velocity on the phase $\varphi$, i.e.
$v_0 \approx \alpha \, (\varphi - \pi)$ such that
the
phase can be determined from the velocity of the ``dark notch''.

Other extreme regimes, as the Tonks-Girardeau gas of Bosons or
an ideal Fermi gas 
diminish the interference, except, in the later case, when an auxiliary
parity-selection procedure is applied to retain odd-states.
A peak is then formed     
with linear dependence on the imprinted phase, very stable 
with respect to the smoothness of the profile of the imprinting laser.    

Finally, note that even though there are no atom-atom interactions
in the reference case, some 
of the aspects usually attributed to the solitons may already be recognized, 
in particular the formation of the dark notch in momentum representation and
its shift with the value of the phase jump or its smoothness. 
Dynamical studies of the state evolution will provide further comparison with soliton
dynamics and will be dealt with elsewhere.

\section*{Acknowledgments}

We acknowledge fruitful conversations with M. Siercke,
C. Ellenor, and A. Steinberg. We further acknowledge
``Acciones Integradas'' of the German
Academic Exchange Service (DAAD) and Ministerio de
Educaci\'on y Ciencia,   
and additional support 
from the Max Planck Institute
for the Physics of Complex Systems;
MEC (FIS2006-10268-C03-01); and UPV-EHU (GIU07/40).  
AR acknowledges support by the German Research Foundation (DFG) and
the Joint Optical Metrology Center (JOMC, Braunschweig).

\end{document}